\documentstyle[preprint,aps,prl,amsmath,amssymb]{revtex}
\begin{document}
\title{ Scattering Theory of Noise:
Current and Charge Fluctuations in
an ideal Conductor.}
\author{G.B. Lesovik}

\address{L.D. Landau Institute of Theoretical Physics, Russian Academy of
Science, \ \\
Kosygina str.2, 117940 Moscow,Russia.}
\maketitle
\ \\
\begin{abstract}
We present the most complete scattering theory for noise
in noninteracting case.
The exact formula for spectral density
of current fluctuations  at finite frequency is presented in terms of
scattering matrix for a coherent quantum conductor.
 We show that the shot noise at finite frequency in an ideally
conducting contact is finite and the spectral density of the noise
 has a singularity at the "Josephson"
frequency $\omega =eV/\hbar $, set by the voltage.
We also discuss fluctuations of charge in an ideal conductor.

The manuscript is a revised and corrected version of JETP Lett.,{\bf 70}, 208 (1999).
The discussion about charge fluctuations is restored (dropped from
the JETPL text due to limitation in space).
\end{abstract}

 In an ideal conductor with electron transmission  equal to unity ($T=1$)
the low-frequency shot noise is suppressed. This has been shown theoretically,
using  quantum-mechanical description in terms of the scattering matrix, for
a quantum point contact \cite{lesovik,yurke,buttiker,martin},
and even earlier,
using kinetic equation for Fermions \cite{kulik} and
 semi-classical Green's functions description \cite{khlus},
for ballistic point contacts.
It has been also confirmed experimentally \cite{reznikov,glattli}
for quantum point contacts.
The same phenomenon has been predicted theoretically \cite{kane}
and confirmed experimentally \cite{glattli2,reznikov2}
for conductors in the fractional quantum Hall
effect regime.
Nonetheless, at finite frequency the noise remains {\it finite } even
in an ideal quantum conductor, and it is this effect that is discussed in the
present letter.

 In a conductor where electron transport can be adequately described in a
model of noninteracting electrons using the scattering matrix, the spectral
density of the shot noise at zero frequency
is given by \cite{lesovik,yurke,buttiker,martin},
\begin{equation}
S(0)=2e^3V/h\sum_nT_n(1-T_n),
\end{equation}
where $V$ is the bias, $e$ is the charge of electron, $ T_n $ is the
transmission in the eigenchannel n, $h$ is Planck's constant, and the
temperature $\Theta $  is assumed to be zero. From this expression it follows
directly that shot noise is completely suppressed for $T=1.$

 The behavior of shot noise only with $T=1$ at low frequencies has been
discussed qualitatively many times, and we shall repeat only briefly
the corresponding arguments. The first condition for the existence of shot
noise is discreteness of the electron charge. Indeed,
if the charge is set equal to zero at a fixed current,
then the noise vanishes, since from (1) it follows that $S(0)/I\propto e$.

 At zero temperature and finite voltage the only reason for fluctuations
of current (besides the already mentioned discreteness of the charge)
transported by noninteracting electrons is the probabilistic nature of
tunneling through a barrier (or scattering by impurities), more accurately,
the probabilistic nature of the measurement process in which the electrons
are detected either in left-hand reservoir (electrons which have been
reflected from a barrier) or in right-hand reservoir (electrons which have
passed through a barrier).
Therefore, if the transmission $T$ becomes unity, this reason for fluctuations
disappears.
The absence of low-frequency fluctuations
due to the irregularity of the electron injection from a reservoir
is due to the fact that electrons obey Fermi statistics.

 We shall now discuss noise at a finite frequency. To understand why noise
is finite in this case, we must refine the arguments presented above. Noise
at zero frequency could be related to fluctuations of the charge
transported through the conductor in time $t$ as follows
\cite{schottky}:
$\lim_{t\rightarrow \infty } \langle (\delta Q_t)^2 \rangle /t=S(0).$

For this reason, if fluctuations of the transported charge grow slower
than $t$, noise at zero frequency is zero. This is what happens for a
conductor with $T=1.$

At the same time, even in an ideal conductor at zero temperature
and finite voltage transmitted charge fluctuations are not zero:

$\langle (\delta Q_t)^2 \rangle = 2e^2/\pi^2 \ln{t \Omega}+
e^2/4\pi^2[(eV/2\epsilon_F)^2-\sin^2(eVt/2\hbar)(\hbar/\epsilon_Ft)^2],$

where $\Omega $  is a cutoff frequency and $\epsilon_F$ is the Fermi energy.
The first term in
this expression is related with the existence of zero-point fluctuations
\cite{levles} and,
in turn, with the fact that the Fermi correlations cannot completely suppress
fluctuations and fix the position of electrons relative to one another. The
second term depends on the applied voltage and is (conventionally speaking) a
result of fluctuations which are due to the transitions of left-moving
electrons into right-moving  electrons and vice versa.

 The finiteness of the noise at finite frequency can now be argued as follows.
Although for $T=1$ a wave packet leaving one reservoir completely passes
into the
other reservoir and the probability of detecting an electron there in time
approaches unity, the charge transport process is nonetheless not completely
uniform in time, and oscillations occur at the frequency of the applied
voltage. Specifically, for a symmetrized current current correlation function
the main (at long times $t\gg \hbar/eV $ )
contribution to the oscillatory dependence has the form

$\langle I(0)I(t)\rangle=\frac{e^2}{\pi^2}
\frac{(eV)^3}{8\epsilon_F^2\hbar t}\sin(eVt/\hbar).$

 Technically, the existence of such fluctuations is due to the fact that
the matrix element of the current operator between two plane waves with
opposite signs of wave vectors,

$\langle \exp(ik_1x)|\hat I|\exp(-ik_2x)\rangle =
\frac{ie\hbar}{2m} [-ik_1+ik_2] \exp(-[ik_1+ik_2]x),$

is nonzero if the absolute magnitudes of the wave vectors are not
equal to one another.
This gives rise to transitions between left-moving and right-moving
electrons. At zero frequency such transitions do not contribute to
single-particle quantities, such as conductivity, but they do appear in
the two-particle correlation function, specifically, in the current correlation
function. Note, that in the linear spectrum approximation,
the effect vanishes.

 We shall now calculate the total current correlation function whence follow
the expressions presented above. The description of transport in a
nonequilibrium system can be given most systematically using Keldysh's
approach \cite{keldish}.

 In this approach to transport, just as with the use of kinetic
equation, the deviation from equilibrium caused
by the presence of an electromotive force in the circuit
is described by prescribing boundary conditions on the
"Keldysh" or "$-+$" Green's function
$G_{-+}(x_1,x_2)=i\langle\Psi^{\dagger}(x_2)\Psi(x_1)\rangle $
at the edges:
\begin{equation}
G_{-+}(x_1,x_2)|_{x_n=X_{\beta}}=G_{-+}^{eq}(X_{\beta},X_{\beta})
\end{equation}

Here  $G_{-+}^{eq}(X_{\beta},X_{\beta})$
is the equilibrium Green's function in the reservoir $\beta $. It can be
shown that for noninteracting system of electrons this approach is completely
equivalent to the approach developed by Landauer and others
\cite{landauer}. In Landauer's
approach it is assumed that the electrons leaving different reservoirs are
uncorrelated. The occupation numbers for the corresponding states are
determined by the chemical potential and the temperature in the corresponding
reservoir. Electron scattering by a static potential inside the conductor
connecting the reservoirs can be described purely quantum mechanically using
wave functions which are exact solutions of the scattering problem.
 What we have said above can be conveniently described in the
second-quantization formalism.

The electronic  $\psi $ - operators are written in a
basis of Lippman-Schwinger scattering states
$\hat \Psi (r)=
\sum_{\alpha,\epsilon,n}
\hat c_{\alpha,\epsilon,n}\varphi_{\alpha,\epsilon,n}(r)$ which have
(in the reservoir $\beta$) the asymptotic form
\begin {equation}
\varphi_{\alpha ,\beta ,\epsilon ,n}(r)=
\frac{\exp(-ik_{nn}|x|)}{\sqrt{v_{nn}}}
\chi_{\beta ,\epsilon ,n,-}(y)\delta_{\beta ,\alpha}+
\sum_mS_{\alpha ,\beta ,\epsilon,n,m}
\frac{\exp(+ik_{nm}|x|)}{\sqrt{v_{nm}}}
\chi_{\beta,\epsilon,m,+}(y).
\end {equation}

The creation operators
$\hat c^{\dag }_{\alpha,\epsilon,n}$ create states of electrons
emitted from reservoir $\alpha $ with energy $E$ in the transverse
channel $n$.
To make the expression less heavy, the conductor is
assumed to be two-dimensional.
The functions $\chi_{\beta,\epsilon,m,\pm }(y)$
describe motion along the $y$ axis.
In the presence of magnetic field these functions describe two different
classes of the edge states (left-moving and right-moving),
and  we introduce indices $\pm $ for these two classes.
It is necessary, because the wave vectors are chosen
to be positive. The magnitudes of the wave vectors are defined by the equality
$\hbar^2k_{nm}^2+ \epsilon_{m}=\epsilon,$
where  $\epsilon_{m}$ is the quantization energy in the $m$th
transverse mode (channel).

 The density matrix of the system can be written, in accordance with the
assumption that the reservoirs are uncorrelated, as a direct product of
independent factors describing each reservoir:
\begin{equation}
\hat \rho =\Pi_{\alpha} \exp(- \sum_{\epsilon,n}
\hat c^{\dag }_{\alpha,\epsilon,n}\hat c_{\alpha,\epsilon,n}
(\epsilon -\mu_{\alpha})]/\theta_{\alpha})/{\bf Tr}\hat \rho .
\end{equation}
where
$\mu_{\alpha}$ and $\theta_{\alpha}$ are the chemical potential
and temperature in reservoir $\alpha .$

 Now using Eqs. (3)-(4) to calculate the correlation
 function of the total currents in the
reservoirs $\beta^{\prime}$  and $\beta $
in the sections $x_1$  and $x_2$  at finite frequency,
\begin {eqnarray}
\langle \langle I_{-\omega}(\beta^{\prime},x_1)
I_{\omega}(\beta ,x_2)\rangle \rangle =
\int dt\exp(i\omega t)
{\bf Tr}\{\hat \rho \hat I_{\beta^{\prime}}(x_1)
e^{i\hat Ht}\hat I_{\beta}(x_2)e^{-i\hat Ht}\}- \\  \nonumber
\langle \hat I_{\beta^{\prime}}(x_1)\rangle\langle\hat I_{\beta}(x_2)\rangle,
\end {eqnarray}
(here $\hat H=\sum_{\alpha,\epsilon,n}
\hat c^{\dag }_{\alpha,\epsilon,n}\hat c_{\alpha,\epsilon,n}\epsilon $),
we obtain the expression
\begin{eqnarray}
\langle \langle I_{-\omega}(\beta^{\prime},x_1)
I_{\omega}(\beta ,x_2)\rangle \rangle =
2e^2\sum_{\alpha^{\prime }n^{\prime }\alpha n}
\int\frac{d\epsilon }{h}
n_{\alpha^{\prime }}(\epsilon^{\prime })[1-n_{\alpha }(\epsilon)]
J^{\beta^{\prime}\alpha^{\prime }\epsilon^{\prime}n^{\prime
}}_{\beta^{\prime}\alpha\epsilon n}
(x_1)
J^{\beta\alpha\epsilon n}_{\beta\alpha^{\prime}\epsilon^{\prime} n^{\prime}}
(x_2)  \  \   .
\end{eqnarray}

Here (and in the formulas below)
$\epsilon^{\prime}=\epsilon +\hbar\omega ,
\epsilon^{\prime }=\epsilon (k^{\prime}), $ and Zeeman splitting is
neglected. The index $\beta $  enumerates the reservoirs where the current is
measured and the index $\alpha $  enumerates the reservoirs
from which electrons are injected.
The current matrix elements can be expressed in terms of the
scattering matrix elements
$ S_{\alpha ,\beta ,\epsilon,n,m} $  as follows:
\begin{eqnarray} J^{\beta \alpha^{\prime }
\epsilon^{\prime }n^{\prime }}_{\beta \alpha \epsilon n}=
\sum_{m^{\prime },m}S^*_{\alpha^{\prime } ,\beta ,\epsilon^{\prime},n,m}
S_{\alpha ,\beta ,\epsilon,n,m}\times   \ \ \ \ \  \ \ \ \  \  \ \ \ \ \ \ \ \ \ \
\ \ \ \ \ \ \ \ \ \ \ \ \ \ \ \ \ \ \ \ \ \\ \nonumber
\times \left(\frac{e\hbar(k^{\prime}_{n^{\prime},m^{\prime}}+k_{n,m})}{2M}
\langle m^{\prime }\epsilon^{\prime}_+|m\epsilon _+\rangle
-\frac{e^2}{Mc}
\langle m^{\prime }\epsilon^{\prime}_+|A_x(y)|m\epsilon _+\rangle\right)
\frac{\exp(-i(k^{\prime}_{n^{\prime},m^{\prime}}-k_{n,m})x)}
{\sqrt{v^{\prime}_{n^{\prime}m^{\prime}}v_{nm}}}+  \ \ \ \ \ \ \ \ \\  \nonumber
\left(\frac{e\hbar(-k^{\prime}_{n,n}-k_{n,n})}{2M}
\langle n\epsilon^{\prime}_-|n\epsilon _-\rangle
-\frac{e^2}{Mc}
\langle n\epsilon^{\prime}_-|A_x(y)|n\epsilon _-\rangle\right)
\frac{\exp(i(k^{\prime}_{n,n}-k_{n,n})x)}
{v_{nn}}\delta_{\beta ,\alpha^{\prime}}\delta_{\beta ,\alpha} + \ \ \ \ \ \\ \nonumber
S^*_{\alpha^{\prime } ,\beta ,\epsilon^{\prime},n^{\prime},m^{\prime}}
\left(\frac{e\hbar(k^{\prime}_{n^{\prime},m^{\prime}}-k_{n,n})}{2M}
\langle m^{\prime }\epsilon^{\prime}_+|n\epsilon _-\rangle
-\frac{e^2}{Mc}
\langle m^{\prime }\epsilon^{\prime}_+|A_x(y)|n\epsilon _-\rangle\right)\times \\ \nonumber
\times \frac{\exp(-i(k^{\prime}_{n^{\prime},m^{\prime}}+k_{n,n})x)}
{\sqrt{v^{\prime}_{n^{\prime}m^{\prime}}v_{nn}}}\delta_{\beta ,\alpha }+ \\ \nonumber
S_{\alpha ,\beta ,\epsilon,n,m}
\left(\frac{e\hbar(-k^{\prime}_{n^{\prime},n^{\prime}}+k_{n,m})}{2M}
\langle n^{\prime }\epsilon^{\prime}_-|m\epsilon _+\rangle
-\frac{e^2}{Mc}
\langle n^{\prime }\epsilon^{\prime}_-|A_x(y)|m\epsilon _+\rangle\right) \times \\ \nonumber
\times \frac{\exp(i(k^{\prime}_{n^{\prime},n^{\prime}}+k_{n,m})x)}
{\sqrt{v^{\prime}_{n^{\prime}n^{\prime}}v_{nm}}}\delta_{\beta,\alpha^{\prime}}  \  ,
\end{eqnarray}
here $v= \partial \epsilon(k)/\hbar \partial k ,$  $x=|x|,$ and
$k$-vectors at the given energy are taken in the lead $\beta .$
Note, that the positive direction of motion is from the origin (scattering region) to
the infinity in each lead, like in polar co-ordinates system.

 The last two terms in the expressions presented above are zero at
zero frequency, and these are the terms that give finite noise at finite
frequency in an ideal conductor, where the scattering matrix has the trivial
form (for two reservoirs denoted by the indices $L$ and $R$)
$|S_{\alpha ,\beta ,\epsilon,n,m}|^2=
\delta_{L,\alpha}\delta_{R,\beta}\delta_{n,m}+
\delta_{R,\alpha}\delta_{L,\beta}\delta_{n,m}.$

 In the presence of a magnetic field the overlap of the transverse wave
functions is
$\langle n^{\prime }\epsilon^{\prime}_+|m\epsilon _-\rangle =
\int dy \chi_{\epsilon^{\prime},n^{\prime},+}^*(y)\chi_{\epsilon ,n,-}(y)$
, and the matrix element
$\langle n^{\prime }\epsilon^{\prime}_+|A_x(y)|m\epsilon_-\rangle) =
\int dy\chi_{\epsilon^{\prime},n^{\prime},+}^*(y)A(y)\chi_{\epsilon ,n,-}(y)$
depends on the energies and the signs of the $k$ vectors ($\pm$).

The contributions
of interest to us, which are due to the nonlinearity of the spectrum, contain
matrix elements of the transverse wave functions describing the edge states
on two different edges. If the width of the conductor $d$ is much greater
than the effective magnetic length,
$d\gg \tilde a_H,$ then noise in an ideal conductor at finite
frequency is exponentially suppressed.



 Without a magnetic field
$\langle m^{\prime}|m \rangle =\delta_{m^{\prime},m}.$
In this case for single channel in the same lead ("right" one)   we obtain

\begin{eqnarray}
\langle \langle \hat I_{-\omega }(x_1)\hat I_{\omega }(x_2)\rangle \rangle=
\frac {2e^2\hbar ^2}{(2m)^2}\int\frac{d\epsilon}{v_{\epsilon^{\prime}}v_{\epsilon}h} \\ \nonumber
\{n_L(\epsilon^{\prime})(1-n_L(\epsilon))
(k+k^{\prime})^2T_{\epsilon^{\prime}}T_{\epsilon}e^{i(k-k^{\prime})(x_1-x_2)} \\ \nonumber
+ n_L(\epsilon^{\prime})(1-n_R(\epsilon))
[(k+k^{\prime})^2T_{\epsilon^{\prime}}(1-T_{\epsilon})e^{i(k-k^{\prime})(x_1-x_2)}+ \\ \nonumber
 (k^{\prime 2}-k^2)T_{\epsilon^{\prime}}e^{-ik^{\prime}(x_1-x_2)}
(r_{\epsilon}e^{ik(x_1+x_2)}+ r^*_{\epsilon}e^{-ik(x_1+x_2)})+
 (k^{\prime}-k)^2T_{\epsilon^{\prime}}e^{-i(k+k^{\prime})(x_1-x_2)}] \\  \nonumber
 n_R(\epsilon^{\prime})(1-n_L(\epsilon))
[(k+k^{\prime})^2T_{\epsilon}(1-T_{\epsilon^{\prime}})
e^{i(k-k^{\prime})(x_1-x_2)}+ \\   \nonumber
 (k^2-k^{\prime 2})T_{\epsilon}
e^{ik(x_1-x_2)}(r_{\epsilon^{\prime}}e^{ik^{\prime}(x_1+x_2)}+
 r^*_{\epsilon^{\prime}}e^{-ik^{\prime}(x_1+x_2)})+
 (k^{\prime}-k)^2T_{\epsilon}e^{i(k+k^{\prime})(x_1-x_2)}] \\  \nonumber
n_R(\epsilon^{\prime})(1-n_R(\epsilon))
 [(k+k^{\prime})^2 e^{-i(k-k^{\prime})(x_1-x_2)}
+(k+k^{\prime})^2R_{\epsilon^{\prime}}R_{\epsilon}
e^{i(k-k^{\prime})(x_1-x_2)}- \\ \nonumber
 -(k+k^{\prime})^2
[r^*_{\epsilon^{\prime}}r_{\epsilon}e^{i(k-k^{\prime})(x_1+x_2)}
+r^*_{\epsilon}r_{\epsilon^{\prime}}e^{-i(k-k^{\prime})(x_1+x_2)}]+ \\  \nonumber
 (k^2-k^{\prime 2})[r^*_{\epsilon}e^{-i(k-k^{\prime})x_1-i(k+k^{\prime})x_2}
-r_{\epsilon^{\prime}}e^{i(k+k^{\prime})x_2}+
r^*_{\epsilon^{\prime}}e^{i(k-k^{\prime})x_2}e^{-i(k+k^{\prime})x_1}
-r_{\epsilon}e^{i(k+k^{\prime})x_1}]+ \\   \nonumber
(k^2-k^{\prime 2})[r^*_{\epsilon^{\prime}}R_{\epsilon}
e^{i(k-k^{\prime})x_1-i(k+k^{\prime})x_2}-r_{\epsilon}
R_{\epsilon^{\prime}}e^{i(k+k^{\prime})x_2}-  \\ \nonumber
R_{\epsilon^{\prime}}r^*_{\epsilon}e^{-i(k-k^{\prime})x_2-i(k+k^{\prime})x_1}
-R_{\epsilon}r_{\epsilon^{\prime}}e^{i(k+k^{\prime})x_1}]- \\  \nonumber
 (k-k^{\prime})^2[r^*_{\epsilon^{\prime}}e^{-i(k+k^{\prime})x_1}
-r_{\epsilon}e^{i(k+k^{\prime})x_1}]
[r^*_{\epsilon}e^{-i(k+k^{\prime})x_2}-r_{\epsilon^{\prime}}e^{i(k+k^{\prime})x_2}]]\},
\end{eqnarray}

where $R_{\epsilon}=|r_{\epsilon}|^2=1-T_{\epsilon},$ and
$n_{R(L)}$ are the occupation numbers in right (left) reservoir.
From this expression we obtain the estimate
$(1-T)\ll (\hbar \omega )^2/\epsilon^2_F$
for values of the transmissions at which the "unusual" contributions are
substantial at finite frequency.

 We shall now write out an expression for the noise in a quantum point
contact, if one channel is completely opened and the wave functions
can be represented in the quasiclassical form
\begin{eqnarray}
\langle\langle\hat I_{-\omega }(x_1)\hat I_{\omega }(x_2)\rangle\rangle=
\frac{2e^2\hbar ^2}{(2m)^2}\int\frac{d\epsilon}{h}\{
\frac{4\omega^2}{(v^{\prime}_1+v_1)(v^{\prime}_2+v_2)\sqrt{v^{\prime}_1v^{\prime
}_2v_1v_2}}  \\ \nonumber
\{n_L(\epsilon^{\prime
})(1-n_R(\epsilon))\exp(-i\int_{x_1}^{x_2}(k^{\prime}(x)+k(x))dx) +\\  \nonumber
 n_R(\epsilon^{\prime
})(1-n_L(\epsilon))\exp(i\int_{x_1}^{x_2}(k^{\prime}(x)+k(x))dx)\} + \\
\nonumber
\frac{(k^{\prime}_1+k_1)(k^{\prime}_2+k_2)}{\sqrt{v^{\prime}_1v^{\prime}_2v_1v_2
}}
\{n_L(\epsilon^{\prime
})(1-n_L(\epsilon))(\exp(-i\int_{x_1}^{x_2}(k^{\prime}(x)-k(x))dx) \\  \nonumber
n_R(\epsilon^{\prime
})(1-n_R(\epsilon))\exp(i\int_{x_1}^{x_2}(k^{\prime}(x)-k(x))dx)\}\}.
\end{eqnarray}

For the part giving shot noise at coinciding co-ordinates we have

\begin{eqnarray}
\langle\langle\hat I_{-\omega }(x)\hat I_{\omega }(x)\rangle\rangle=
\\ \nonumber
\frac {e^2\hbar ^2}{(2m)^2}\int \frac{d\epsilon}{\hbar 2\pi}
\frac{4\omega^2[n_L(\epsilon^{\prime })(1-n_R(\epsilon))+
n_R(\epsilon^{\prime })(1-n_L(\epsilon))]}
{(v_{k^{\prime}}(x)+v_k(x))(v_{k^{\prime}}(x)+v_k(x)v_{k^{\prime}}
(x)v_k(x)}
\end{eqnarray}
It is evident from this expression that the fluctuations are big near a
constriction, where the kinetic energy of the longitudinal motion is
minimal, and they decrease into the reservoir.

 The frequency dependence has a singularity at
$\hbar \omega=eV,$ similar to that studied
previously in a normal contact with scattering
\cite{yang,leslev}
and in a contact between
Luttinger liquids
\cite{chamon}.

For a symmetrized correlation function at voltages much
lower than the minimum of the effective Fermi energy
$\epsilon_F(x)=\epsilon_F-\epsilon_1(x),$ the spectral density has the form
\begin{eqnarray}
\langle\langle\frac{1}{2}[\hat I_{-\omega }(x)\hat I_{\omega }(x)
+\hat I_{\omega }(x)\hat I_{-\omega }(x)]\rangle\rangle=
\frac{2e^2}{h}(\hbar \omega+
\frac{(\hbar\omega)^2}{16\epsilon_F^2(x)}f(\hbar \omega)),
\end{eqnarray}
where $f(\hbar \omega )=eV$ for frequencies
$\hbar\omega < eV,$ and  $f(\hbar \omega )= \hbar\omega $
if  $\hbar\omega >eV.$ Such a singularity can by
observed in direct measurements
\cite{schoelkopf1} and in an experiment with an additional
external frequency; see Refs.20 and 23.
 In experiments where the noise is measured at a finite frequency, if the
measurements are performed using a resonant circuit with high $Q$,
 only "positive frequency" makes a contribution, and the excess noise is
 $\frac{e^2(\hbar\omega )^2}{h8\epsilon_F^2(x)}(eV-\hbar\omega )$
\cite{roland}.

Now we turn to a discussion of the spatial dependence
of the obtained correlators. Physically it is clear
that for stationary process the zero-frequency correlator
$\langle I_{-\omega }(x_1)I_{\omega }(x_2)\rangle|_{\omega=0} $
should not depend on the co-ordinates,
because such a dependence would mean charge accumulation in time.
Let us prove this statement formally.

It follows from the Shroedinger equation that the matrix element of
the current operator, taken between two states of the system
with the energies $E$ and  $E^{\prime}=E+\hbar \omega $ obeys the equation
$\partial/\partial x\langle E+\hbar \omega ,\alpha |I(x)|E,\beta \rangle
=-ie\omega
\Psi^*_{E+\hbar \omega ,\alpha }(x) \Psi_{E,\beta }(x),$
and therefore at zero frequency the $x$-dependence
vanishes.

In addition, since we consider stationary process, the density matrix
must be diagonal in the energy representation:
$\langle E'|\hat \rho|E\rangle \propto \delta (E'-E).$
Using these facts, we conclude that the correlator

$\langle I_{-\omega }(x_1)I_{\omega }(x_2)\rangle|_{\omega=0} =
\sum_{\alpha,\beta,E}\langle \alpha,E|\hat \rho|\beta,E\rangle
\langle E,\beta |I(x_1)|E,\gamma\rangle
\langle E, \gamma |I(x_2)|E\alpha \rangle $

at zero frequency also does not depend on the co-ordinates $x_1,x_2$.

One checks, that the correlator (8) indeed loses its co-ordinate
dependence in the zero-frequency limit,
but this limit is nontrivial in fact, as we discuss below.

In the specific contributions in the correlator we are discussing
the functional dependence on the co-ordinates remains strong even
in the zero-frequency limit,
$\sim \exp(ik_Fx)$,
nevertheless the coefficient in front of such terms vanishes as
second (or first) degree of the frequency.
In the remaining "usual" terms in the correlator,
which give finite contribution
at zero frequency  the functional dependence
becomes weak by itself $\sim \exp(i\omega /v_Fx).$

{\bf Charge fluctuations}.
At the same time, in the charge fluctuations all kinds of terms
give comparable contribution. This fact was also noted for normal conductors
in \cite{levinson99},
 and considered for normal metal-superconductor system
in \cite{nsideal}.

Charge fluctuations in a volume between sections  $x_1,$ $x_2$
 can be  expressed via already known current correlators:

$\langle\langle Q^2_{\omega }(x_1,x_2)\rangle\rangle=\frac{1}{\omega^2 }
\langle \langle I_{-\omega }(x_1)I_{\omega }(x_1)
-I_{-\omega }(x_1)I_{\omega }(x_2)
-I_{-\omega }(x_2)I_{\omega }(x_1)
+I_{-\omega }(x_2)I_{\omega }(x_2)\rangle \rangle ,$

At frequencies much less than minimal Fermi energy
$\min{\{mv_F^2(x)/2\}}$ for the open channel
 we obtain for the charge fluctuations in the volume $[x_1,x_2]$:

\begin{gather}
\langle\langle Q^2_{\omega }(x_1,x_2)\rangle\rangle=
\frac{2e^2}{h}\int d\epsilon
\{[n_L(\epsilon^{\prime })(1-n_L(\epsilon))+
n_R(\epsilon^{\prime })(1-n_R(\epsilon))]
\frac{4\sin^2(\omega \tau/2)}{\omega^2}   \\ \nonumber
+ [n_L(\epsilon^{\prime })(1-n_R(\epsilon))+
n_R(\epsilon^{\prime })(1-n_L(\epsilon))]
\frac{\hbar^2}{4m^2v_1^4v_2^4}\times\\ \nonumber \times
[(v_1^2-v_2^2)^2+4v_1^2v_2^2\sin^2(\int_{x_1}^{x_2}k(x)dx+\omega\tau/2)]\},
\end{gather}
where  $v_i=v_F(x_i),$ $\tau =\int_{x_1}^{x_2}dx/v_F(x)$.

In the present article we consider strictly noninteracting picture,
but we expect the main statement about the finiteness of
current and charge
fluctuations in an ideal conductor to be correct also in presence
of interaction.
It is clear, nevertheless, that an interaction may play important role
in the considered effect. For example Coulomb interaction will
 be important for "transmission"
of the fluctuations through wires into  measuring devises
 \cite{levitov}.

Besides, in the presence of interaction an additional singularities
may appear at frequencies which are multiple of the inverted
traveling time of one electron  $\omega =n 2\pi I/e, n=1,2,...$
These singularities are in general
different from the one we discussed ( at $\omega =eV/\hbar $), the latter
being due to the interference between states from the reservoirs at
different chemical potentials.

It is also interesting to note, that since it is nonlinear dispersion which is
responsible for a finite noise in an ideal conductor, the effect
will be invisible in the usual Luttinger liquid description.
In that case accounting for the deviation from the linear dispersion
as a term in Hamiltonian, mixing "left" and "right" moving electrons
is needed.

 In closing, we note that in previously presented expressions for noise at
finite frequency \cite{lesovik,yang}
the terms making a finite contribution in an ideal conductor
were omitted; the reason was that the corresponding terms are small.
However, the substantial advances made in
resent years in measuring noise and in preparing new mesoscopic
conductors hold
forth the hope that either it will be possible to measure even
weaker noise in
existing structures or that conductors with a very low Fermi energy and high
mobility, where the effect is large, will become available.

As about observing the charge fluctuations, corresponding to
such small terms in current fluctuations in ideal conductor,
situation is yet more promising. As we showed in Expr.12
(see also \cite{levinson99}), these fluctuations are not small, and can be
detected for example in "Which-Path Detector" (WPD) setup \cite{buks},
provided
the relevant length of the part of the wire, in which fluctuations
are probed  by WPD, is smaller or comparable to the effective
Fermi wave-length.

 I thank D. C. Glattli, M. Reznikov, D. Prober, and  especially
H. Bouchiat for helpful and stimulating discussions.


\begin{thebibliography}{99}

\bibitem{lesovik} G.B. Lesovik, JETP Lett. {\bf 49 }, 592 (1989).
\bibitem{yurke} B. Yurke, G.P. Kochanski, Phys.~Rev. B. {\bf 41}, 8184 (1990).
\bibitem{buttiker}M.~Buttiker, Phys. Rev.~Lett.{\bf 65}, 2901 (1990).
\bibitem{martin} Th. Martin, R. Landauer, Phys.~Rev. B. {\bf 45}, 1742 (1992).
\bibitem{kulik} I. O. Kulik, A. N. Omel'yanchuk,
Sov. J.Low Temp.Phys.{\bf 10}158 (1984).
\bibitem{khlus} V.A. Khlus, Sov. Phys. JETP {\bf 66}, 1243 (1987).
\bibitem{reznikov}M. Reznikov et al., Phys. Rev. Lett. {\bf 75}, 3340 (1995).
\bibitem{glattli} A. Kumar et al., Phys. Rev. Lett. {\bf 76}, 2778 (1996).
\bibitem{kane} G.B. Lesovik, A.H. MacDonald, 1993, unpublished,
see remark at Ref. 14 ;C.L. Kane, M.P.A. Fisher ,
Phys. Rev. Lett. {\bf 72}, 724 (1994).
\bibitem{glattli2}  L. Saminadayar et al., Phys. Rev. Lett. {\bf 79}, 2526
(1997).
\bibitem{reznikov2} de-Picciotto et al., Nature {\bf 389}, 162 (1997).
\bibitem{schottky} W. Schottky, Ann. Phys. (Leipzig) {\bf 57}, 16432 (1918).
\bibitem{levles} G. B. Lesovik, L. Levitov,
JETP Lett.,{\bf 58}, 230 (1993).
\bibitem{roland2} G. B. Lesovik, R. Loosen, Z. Phys. {\bf B 91}, 531 (1993).
\bibitem{keldish} L.V. Keldysh, Sov. Phys. JETP {\bf 20}, 1018 (1965).
\bibitem{landauer} R. Landauer, IBM J. Res. Dev.{\bf 1}, 223 (1957);
Phil.~Mag., {\bf 21}, 863 (1970);
P. W. Anderson, D. J. Thouless, E. Abrahams, D.S. Fisher,
Phys.~Rev. B. {\bf 22}, 3519 (1980);
D. S. Fisher, P. A. Lee, Phys.~Rev. B. {\bf 23}, 6851 (1981);
M. Buttiker, Phys.~Rev. Lett. {\bf 57}, 1761 (1986);
Y. Imry, in: "Directions in Condensed Matter Physics",
G. Grinstein and G. Mazenko, Eds., World Scientific Press,
Singapore, 1986, p. 101.
\bibitem{levinson} Y. B. Levinson, Sov. Phys. JETP, {\bf 68},1257, (1989).
\bibitem{chklo} D. B. Chklovskii et al., Phys.~Rev. B. {\bf 46}, 4026 (1993).
\bibitem{yang} S.-R. Eric Yang, Solid State Commun. {\bf 81}, 375 (1992).
\bibitem{leslev} G. B. Lesovik, L. Levitov,
Phys.~Rev.~Lett. {\bf 72}, 538 (1994).
\bibitem{chamon} C. de C. Chamon et al., Phys.~Rev. B. {\bf 53}, 4033 (1996).
\bibitem{schoelkopf1}R. J. Schoelkopf et al.,
Phys.~Rev.~Lett. {\bf 78}, 3370 (1997).
\bibitem{schoelkopf}R. J. Schoelkopf et al.,
Phys.~Rev.~Lett. {\bf 80}, 2437 (1998).
\bibitem{roland} G. B. Lesovik, R. Loosen, JETP Lett. {\bf 65}, 295 (1997).
\bibitem{levinson99} Y. Levinson, cond-mat/9907055 (1999)
\bibitem{nsideal} Rodolphe Guyon, Thierry Martin, Gordey B.
Lesovik, Tunable decoherence in the vicinity of a
normal metal-superconducting junction, cond-mat/0011244,
Journal-ref: Phys. Rev. B 64, 035315 (2001).
\bibitem{levitov} H. Lee, L. S. Levitov, Phys. Rev. B {\bf 53}, 7383 (1996).
\bibitem{buks} E. Buks {\it et al} Nature {\bf 391}, 871 (1998).

\end{thebibliography}
\end{document}